\documentclass[]{aa}
\usepackage{epsfig}
\usepackage{natbib}
\usepackage{txfonts}
\newcommand{\sub}[1]{_{\rm #1}}
\newcommand{\reference}[1]{}
\def\apj{{ApJ }}
\def\ApJ{{ApJ }}

\def\AandA{{A\&A }}

\newcommand{\degree}{{}^{\circ}}
\newcommand{\changed}{}

\begin{document}

\title{Structure analysis of interstellar clouds: II. Applying the
  $\Delta$-variance method to interstellar turbulence}
\titlerunning{$\Delta$-variance analysis of interstellar turbulence}
\author{V.~Ossenkopf\inst{1,2,3}, M.~Krips\inst{1,4} \and{} J. Stutzki\inst{1}}

\institute{1. Physikalisches Institut der Universit\"at 
zu K\"oln, Z\"ulpicher Stra\ss{}e 77, 50937 K\"oln, Germany
\and
SRON Netherlands Institute for Space Research, P.O. Box 800, 9700 AV 
Groningen, Netherlands
\and
 Kapteyn Astronomical Institute, University of Groningen, PO box 800, 
 9700 AV Groningen, Netherlands
\and
Harvard-Smithsonian Center for Astrophysics, SMA project,
60 Garden Street, MS 78 Cambridge, MA  02138, USA}

\date{Received: December 23, 2002; accepted February 22, 2008}

\abstract
{
The $\Delta$-variance analysis is an efficient tool for measuring
the structural scaling behaviour of interstellar turbulence in
astronomical maps. It has been applied both to simulations of
interstellar turbulence and to observed molecular cloud maps.
In paper I we proposed essential improvements to the 
$\Delta$-variance analysis and tested them on artificial
structures with known characteristics.
}
{
In this paper we apply the improved $\Delta$-variance analysis
to simulations of interstellar turbulence and observations of
molecular clouds. We tested the new capabilities in practical use 
and studied properties of interstellar turbulence
that could not have been addressed before.
}
{
We selected three example data sets that profit in particular
from the improved $\Delta$-variance method: i) a hydrodynamic 
turbulence simulation with prominent density and velocity structures,
ii) an observed intensity map of $\rho$ Oph with irregular boundaries 
and variable uncertainties of the different data points, and iii) a map 
of the turbulent velocity structure in the Polaris Flare affected
by the intensity dependence on the centroid velocity determination.
}
{
The tests confirm the extended capabilities of the improved
$\Delta$-variance analysis. Prominent spatial scales were accurately
identified and artifacts from a variable reliability of the data
were removed.
The analysis of the hydrodynamic simulations showed that
the injection of a turbulent velocity structure creates the most 
prominent density structures are produced on a scale somewhat below 
the injection scale. The new analysis of a $\rho$ Oph continuum map 
reveals an intermediate stage in the molecular
cloud evolution showing both signatures of the typical molecular
cloud scaling behaviour and the formation of condensed cores.
When analysing the velocity structure of the Polaris Flare we show
that a universal power law connects scales from 0.03~pc to
3~pc. However, a plateau in the $\Delta$-variance spectrum
around 5~pc indicates that the visible large-scale velocity gradient
is not converted directly into a turbulent cascade here. It is
obvious that, for any turbulent structure, effects of low-number 
statistics become important on the driving scale.
}
{}
\keywords{Methods: data analysis -- Methods: statistical --
ISM: clouds -- ISM: structure}

\maketitle

\section{Introduction}

{\changed Observations of interstellar clouds show a complex, filamentary structure
which can be attributed to turbulence in the interstellar medium
\citep{Puebla, Roma, MLK}. To understand the processes governing the
structure and evolution of the clouds, turbulence models have to be
constructed and compared to observational data. Their parameters and 
implementational details need to be adjusted to fit the observed 
behaviour. Due to the random nature of turbulence, simulations will
never provide an exact reproduction of the observed data sets but will
only reproduce general statistical properties like scaling relations. 

The most common scaling relation to characterise turbulent structures is
the power spectrum of fluctuations, both in density and in velocity.
Here, we encourage using another quantity, the $\Delta$-variance spectrum,
introduced by \citet{Stutzki}, a wavelet-based method to measure the
relative amount of structural variation as a function of the size scale.
Due to the mutual relations between the $\Delta$-variance spectrum
and the radially averaged power-spectrum, the $\Delta$-variance analysis
can be considered as a very robust method of evaluating the power 
spectrum of a structure. The advantages of the $\Delta$-variance method
result from the smooth wavelet filter shape, which provides
a robust way for an angular average independent of gridding
effects, and from the insensitivity to edge effects as discussed
by \citet{Bensch}.}

In parallel to the structure scaling analysis, clump {\changed 
decomposition algorithms
like GAUSSCLUMP \citep{SG} found that the clump mass spectrum 
$dN/dM \propto M^{-\gamma}$ of sufficiently large molecular cloud
data sets also tend to follow power laws over many orders of magnitude}
with the spectral index $\gamma$ in a relatively narrow 
range between 1.7 to 1.9 \citep{Kramer, Heithausen98}.
\citet{Stutzki} demonstrated that a clump ensemble with such
a mass spectrum and with a power-law mass-size scaling relation 
results in a cloud image
with the $\Delta$-variance scaling index {\changed determined by the
number-mass and the mass-size spectral indices. However, it was
questioned by e.g. \citet{Vazquez-Semadeni, Ballesteros-Paredes} whether
the observed mass-size relation reflects true properties of the
underlying structure. It may rather represent an observational artifact.}
From these results it is obvious that there is a strong interest in studying
the spatial scaling behaviour of observed maps of the interstellar medium
via the $\Delta$-variance analysis.

{\changed
In paper I we have proposed several essential improvements
to the original $\Delta$-variance method. We have investigated the
use of different wavelets and calibrated their spatial resolution.
Unfortunately, it turns out that it is not possible to define a 
single optimum wavelet for all purposes because different wavelets 
exhibit a different power in the detection of the characteristic
structures. A good compromise is given by the Mexican-hat filter
with a diameter ratio of 1.5. When the main focus lies on the
measurement of the spectral index, the French-hat filter with a diameter
ratio of about 2.3 is also suitable. We also introduced
a significance function to weight the different data points.
This allows us to analyse observed data where the signal-to-noise
ratio is not uniform across the mapped area, but spatially varying.
This should permit to distinguish the influence of variable noise from
actual small-scale structure in the maps.
The need for such a treatment became very obvious when 
\citet{OML} used the $\Delta$-variance analysis to characterise the
velocity structure detected in molecular line observations of the
Polaris Flare taken by Falgarone et al. (1998), \citet{Bensch}, and 
\citet{Heithausen90}. Comparing the $\Delta$-variance analysis with
the size-linewidth relation, we found that the $\Delta$-variance of 
the centroid velocity maps produced wrong results on scales where the 
maps do not show any noticeable emission. With the introduction of
the significance function, the new $\Delta$-variance analysis should be
able to also reliably analyse such data sets. Moreover, the use
a weighting function also allows us to use computational methods like 
the fast Fourier transform to obtain the $\Delta$-variance spectrum 
even for maps with irregular boundaries or maps which are only sparsely
filled by significant values.

In paper I we tested the properties of the improved $\Delta$-variance
analysis using simple artificial data sets. In this paper we will
apply it to more realistic data sets, either simulations of
interstellar turbulence or to actual observational data. In Sect. 2
we recapitulate the formalism of the new $\Delta$-variance analysis
and summarise the results that we obtained from the application
to the test data sets. In Sect. 3 we apply the analysis to a
hydrodynamic simulation, to the $\rho$-Oph dust continuum
map by \citet{Motte} and to the centroid velocity map for the Polaris
Flare. We  discuss the conclusions on their structure in Sect. 4.}

\section{The improved $\Delta$-variance method}

In this section we summarise the main properties of the improved
$\Delta$-variance analysis proposed in paper I, focusing on the
new points, not covered in the original $\Delta$-variance
definition by \citet{Stutzki}.

The $\Delta$-variance measures the amount of structure on a given scale
$l$ in a map $f(\vec{r})$ by filtering the map with a spherically symmetric 
wavelet of size $l$ and computing the variance of the thus filtered map:
\begin{equation}
\sigma_\Delta^2(l)= \left\langle \left( f(\vec{r}) * {\bigodot}_l(\vec{r}) \right)^2 \right\rangle_{\vec{r}}
\label{eq_basicdelta}
\end{equation}
where, the average is taken over the area of the map, 
the symbol $*$ stands for a convolution, and $\bigodot_l$ describes
the filter function composed of positive inner ``core''
and a negative annulus, both normalised to integral values of unity
\begin{equation}
{\bigodot}_l(\vec{r}) = \bigodot_{l, {\rm core}}(\vec{r})
- \bigodot_{l, {\rm ann}}(\vec{r}) \;.
\end{equation}
We have studied the ``French-hat'' filter with constant values in both parts:
\begin{eqnarray}
\bigodot_{l, {\rm core}}(\vec{r}) &=& {4 \over \pi l^2}
\left\{ \begin{array}{ll} 1 &: |\vec{r}| \le l/2\\
0 &: |\vec{r}| > l/2 \end{array} \right. \nonumber \\
\bigodot_{l, {\rm ann}}(\vec{r}) &=& {4 \over \pi l^2}
\left\{ \begin{array}{ll} {1/(v^2-1)} &: l/2 < |\vec{r}| \le v \times l/2\\
0 &: |\vec{r}| \le l/2, |\vec{r}| > v\times l/2 \end{array} \right.
\label{fhat}
\end{eqnarray}
and a ``Mexican hat'' consisting of two Gaussian functions:
\begin{eqnarray}
\bigodot_{l, {\rm core}}(\vec{r}) &=& {4 \over \pi l^2}
\exp\left(\vec{r}^2 \over (l/2)^2 \right) \\
\bigodot_{l, {\rm ann}}(\vec{r}) &=& {4 \over \pi l^2 (v^2-1)}
\left[ \exp\left(\vec{r}^2 \over (vl/2)^2\right)
- \exp\left(\vec{r}^2 \over (l/2)^2\right)\right]\nonumber
\end{eqnarray}
where $l$ is the core diameter and $v$ is the diameter ratio
between the annulus and the core of the filter.
Plotting the $\Delta$-variance
as a function of the filter size $l$ then provides a spectrum
showing the relative amount of structure in a given map as a
function of the structure size.

The effective filter size, given by the average distance of points
in the core and the annulus, deviates from the core diameter
$l$ as
\begin{equation} 
{l\sub{eff} \over l} =\left\{\displaystyle
0.29 v+0.26 \quad  {\rm for\; the\; French\; hat} \hfill
\atop \displaystyle
0.41 v+0.46 \quad {\rm for\; the\; Mexican\; hat.} \hfill \right.
\end{equation}
Thus structures with a particular size should show
up as prominent peaks in the $\Delta$-variance spectrum on a scale 
$l\sub{eff}$ corresponding to that size. Test with artificial
data sets in paper I have shown, however, that the peak positions
always falls 10-20\,\% below the maximum structure size. Taking 
this systematic offset into account we can nevertheless reliably
calibrate the spatial resolution of the $\Delta$-variance analysis.

A major improvement of the new $\Delta$-variance algorithm 
was the introduction of a weighting function to the data 
$w\sub{data}(\vec{r})$.
This simultaneously solved the problems of the edge treatment of
finite maps and the analysis of data with a variable uncertainty 
across the map. The weight function varies between 0 and 1, representing
the reliability of the individual data points, and it extends beyond the 
original map size, padding it with zeros at the boundaries. 
Instead of the original map $f(\vec{r})$, an extended map, 
$f\sub{padded}(\vec{r})=f(\vec{r}) \times w\sub{data}(\vec{r})$ inside
the original data area, $f\sub{padded}(\vec{r})=0$ outside, is
analysed. This padded map can be periodically continued without 
wrap-around effects, so that the filter convolution can be efficiently
computed 
in Fourier space involving a fast Fourier transform and a map multiplication.

To avoid that data points within the padded area or with a low
weighting are counted like normal zero-value data, but disregarded 
in the computation of the variance, the filter has to be re-normalised
at each position in the map in such a way that the integral weights
of core and annulus remain unity when excluding the padded points and
when taking the weighting of the normal data points into account. 
Instead of one convolution (Eq.~\ref{eq_basicdelta}), one has
to compute four convolutions
\begin{eqnarray}
G_{l, {\rm core}}(\vec{r}) &=& f\sub{padded}(\vec{r}) * 
\bigodot_{l, {\rm core}}(\vec{r'}) \nonumber \\
G_{l, {\rm ann}}(\vec{r}) &=& f\sub{padded}(\vec{r}) * 
\bigodot_{l, {\rm ann}}(\vec{r'}) \nonumber \\
W_{l, {\rm core}}(\vec{r}) &=& w(\vec{r}) * 
\bigodot_{l, {\rm core}}(\vec{r'}) \nonumber \\
W_{l, {\rm ann}}(\vec{r}) &=& w(\vec{r}) * 
\bigodot_{l, {\rm ann}}(\vec{r'}) 
\end{eqnarray}
and combine the results while re-normalising with the effective
filter weight for the valid data
\begin{equation}
F\sub{l}(\vec{r}) = {G\sub{{\it l}, core}(\vec{r}) \over
W\sub{{\it l}, core}(\vec{r})}
 - {G\sub{{\it l}, ann}(\vec{r}) \over
W\sub{{\it l}, ann}(\vec{r})} \;.
\end{equation}

From the actual filter weight computed for each point in the map we can derive
a significance function as the product of both normalisation factors
\begin{equation}
W\sub{{\it l}, tot}(\vec{r})=W\sub{{\it l}, core}(\vec{r})
W\sub{{\it l}, ann}(\vec{r})\;.
\end{equation}
This provides the actual significance of the data points in the
convolved map which is used when computing the $\Delta$-variance 
of the whole map 
\begin{equation}
\sigma_\Delta^2(l) = { \sum\sub{map} (F\sub{l}(\vec{r}) - \langle F\sub{l}
\rangle )^2 W\sub{{\it l}, tot}(\vec{r}) \over \sum\sub{map}
W\sub{{\it l}, tot}(\vec{r}) }\;.
\end{equation}

With this generalised concept, the $\Delta$-variance analysis can be applied
to arbitrary data sets. They must be 
projected onto some regular grid but they do not need to contain regular
boundaries as the corresponding ``empty'' grid points can be marked with 
a zero significance. Varying noise or other changes
in the data reliability can be expressed in the significance
function $w\sub{data}(\vec{r})$. This applies e.g. to maps
where not all points are observed with the same integration time
so that they show a different noise level. In paper I we used
a weighting function given by the inverse noise RMS and in 
Sect. \ref{sect_andre} we will study the impact of the selection
of the weighting function for observed data.

The only remaining requirement for the applicability of the
$\Delta$-variance analysis is a sufficiently large spatial dynamic range
in the data. \citet{Bensch} had shown that a map has to contain
at least 30 pixels in each direction to obtain reasonable error 
bars of the $\Delta$-variance spectrum. Numerical tests with noisy
data in paper I showed that this critical size needs to be extended by
a factor of about one over the average data significance in the case
of data with a variable reliability.

\section{Applications}
\label{sect_applic}

\subsection{Hydrodynamic simulations}
\label{sect_mhd}

In the papers by \citet{MLO}, \citet{OML}, and \citet{OKH} we
have demonstrated the general applicability of the $\Delta$-variance
analysis to extract characteristic structure sizes and scaling
laws from (magneto-)hydrodynamic simulations performed with a
variety of codes. Here, we need to test whether the $\Delta$-variance
with adapted filter functions improves the sensitivity of this
method. The weighting function is irrelevant in this case 
because the data do not suffer from noise or another cause
of variable reliability across the data set.

We have applied the analysis to a variety of simulations presented
in \citet{OKH}, but we present the results here only for a single
model, the first inertial stage of the small-scale driven hydrodynamic
turbulence computed by smooth-particle hydrodynamics (SPH), S02 at
$t=0$. In this simulation the velocity field is driven by a Gaussian
field of random fluctuations within a finite wavenumber range, $k=7\dots8$.
This means, that the driving process introduces characteristic variations
into the velocity structure with the same scale length as used
in the artificial sine wave field used in paper I,
but with wavenumbers between 7 and 8. Thus the $\Delta$-variance spectrum
should measure a peak variation for scales of $1/(\sqrt{2} k)$, i.e.
between 0.088 and 0.101 of the size of the whole data cube. 
Selecting a model which is driven on small scales guarantees
that we can identify a clear peak for these structures leaving enough
dynamic range on smaller and larger scales. We select the initial
stage of fully evolved turbulence in the simulation to make sure that
the turbulent driving is the only process creating structures in the 
data set, avoiding effects of self-gravity. In this way we have a data
set which is best suited to test the structure recognition by the 
$\Delta$-variance analysis, as it should directly detect the scales
of the driving process in the velocity structure and determine the
scaling of the turbulent cascade on lower scales. Results for the
other simulations did not provide any fundamentally different results, 
but are less clear to interpret because either the 
driving scale is closer to the edge of the dynamic range or the
contained structure is less well known.

\begin{figure}
\epsfig{file=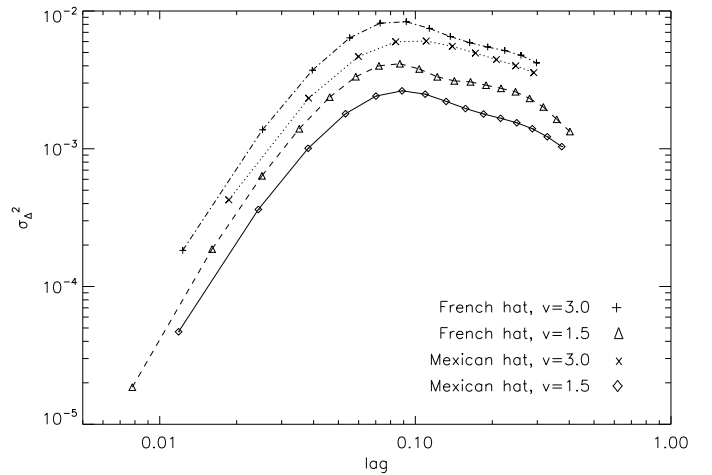, angle=90, width=\columnwidth}
\caption{$\Delta$-variance spectra of the $z$-projection of
the $z$-component of the velocity field in the hydrodynamic
simulation S02 at $t=0$. Four different filter shapes are tested against
each other.
} 
\label{fig_hdzvel}
\end{figure}

Figure~\ref{fig_hdzvel} shows the $\Delta$-variance spectra of the $z$-projection
of the $z$ component of the velocity field of the turbulence simulation
computed with four different filter functions. We find a clear shift between the
peak positions measured by the different filters. The Mexican-hat filter
gives systematically larger lags for the peak than the French hat and the
lag of the peak grows with growing annulus-to-core diameter ratio, $v$.
The minimum peak lag, given by the French hat with $v=1.5$, falls at 0.084,
the maximum lag, given by the Mexican hat with $v=3$, at 0.11. 

This behaviour
is consistent with the results obtained for the simple sine wave field
in paper I. It can be understood as a result of the
variable width and the shape of the filters in Fourier space. The
broadening of the filter function reduces the contrast which leads in 
a spectrum with a steep slope on small scales and a shallow slope on large
scales to an effective shift of the peak position. The steep decay of
the French-hat filter function for large lags leads to a somewhat lower
peak position, but is always accompanied by side lobes of the Bessel function
visible as artificial secondary peaks at large lags. For the turbulence
simulations these secondary peaks are not as pronounced as for the 
sine wave field, but they are also visible in Fig.~\ref{fig_hdzvel}.
For the diameter ratios $v$ of about 1.5 for the Mexican hat and
2.3 for the French hat, deduced as optimum values in paper I, the
peak position falls at about 0.088 in both cases, a value slightly lower
than the expected average structure size.

The different width of the peak has a direct impact on the slope of
the turbulent structures measured at small lags. With the broad peaks
produced by the Mexican-hat filter, the slope is affected down to
relatively small scales. As the filter diameter ratio $v$ constrains
the minimum scale which can be resolved, a clear power law becomes
only visible for the French-hat filter with $v=1.5$. 
Small diameter ratios are always favourable with respect to the
dynamic range which can be covered in the $\Delta$-variance analysis
because of the minimum filter size and the constraint that the overall
filter must always remain small compared to the analysed map.
For the French hat 
with $v=3.0$ and the Mexican hat with $v=1.5$ the remaining dynamic range
is just marginally sufficient to reliably determine the spectral index.
We find slopes between 2.8 and 3.1 corresponding to power spectral
indices $\zeta=4.8\dots 5.1$. This is much higher than the Kolmogorov
index of $\zeta=3.67$ indicating that the scaling on small
scales is not determined by a self-similar turbulent
cascade, but by the numerical viscosity in these simulations 
damping small scale structures. For simulations driven on large scales
resulting in a larger dynamic range below the peak, we find a limited
inertial range with a slope of about two, corresponding to $\zeta \approx 4$
which is consistent with a cascade of Burger's turbulence 
\citep[see]{OML,OKH}.

\begin{figure}
\epsfig{file=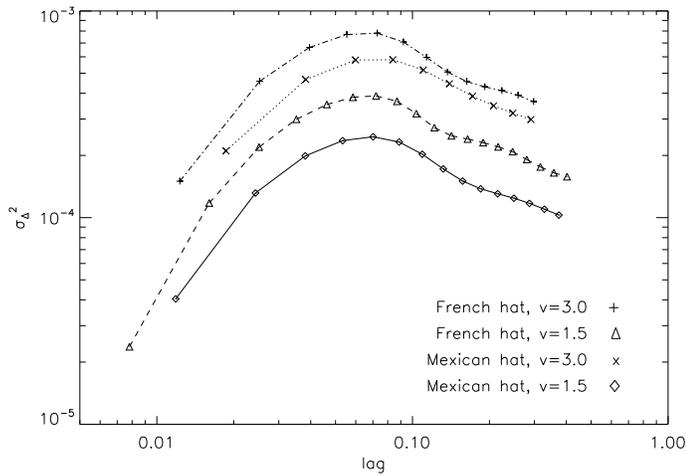, angle=90, width=\columnwidth}
\caption{$\Delta$-variance spectra of the density structure
created by the velocity field from Fig.\ref{fig_hdzvel}, measured
with four different filter functions.
} 
\label{fig_hddens}
\end{figure}

In a second step we investigate how the turbulent velocity scaling
translates into the creation of turbulent density enhancements.
Figure~\ref{fig_hddens} shows the $\Delta$-variance spectra of 
the density structure seen in the same step of the simulation
using the same filters as applied to the velocity structure. We find
the same systematic deviations between the results seen by
different filters but a generic shift of the peak position to
shorter lags by a factor 0.75-0.8 with respect to the peaks of the
velocity structure. The slope of the turbulent density structure on
small scales is shallower by $1.0\pm 0.05$. For the mutual comparison
between density and velocity structure, the selection of the filter
is thus irrelevant as long as the same filter is applied in both
cases.

We see that injesting energy on a particular scale does not create
density enhancements on that scale, but rather on a scale smaller
by a factor 0.75-0.8. This shift has not been noticed before by
\citet{MLO} as only the systematic tests of the filter functions 
provided enough sensitivity with respect to a reliable scale detection.
It seems that the turbulent cascade builds up density fluctuations
on all scales below the driving scale, but that those density
enhancements act themselves as points of an efficient 
energy conversion between the scales creating new density structures
so that dominant density scale falls somewhat below the
initial scale.

\begin{figure}
\epsfig{file=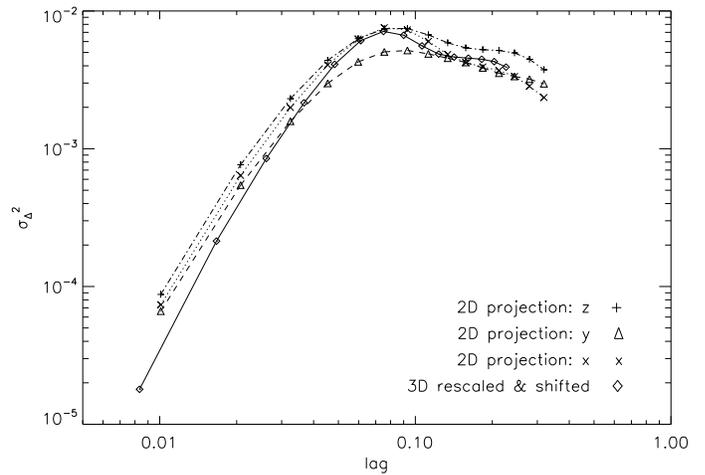, angle=90, width=\columnwidth}
\caption{$\Delta$-variance spectra of the different rectangular projection
maps and of the full three-dimensional data cube of
the $z$-component of the velocity field in simulation S02 at $t=0$. The
analysis used the French-hat filter with a annulus-to-core diameter
ratio $v=2.3$
} 
\label{fig_hdvelcomponents}
\end{figure}

Finally we need to address the significance of the measured
structure size and the scaling indices with respect to the random
fluctuations always present in turbulence simulations and with respect
to the relation between the three-dimensional (3-D) structure and the 
two-dimensional (2-D) projections which we can measure in astronomical 
observations. For turbulent density structures this comparison has been
done by \citet{MLO} using the ``traditional´´ $\Delta$-variance filter 
with the diameter ratio $v=3.0$. We will repeat it here for the velocity
structure as the direct carrier of turbulent energy using one of
more sensitive filter functions.

Figure \ref{fig_hdvelcomponents} shows the $\Delta$-variance spectra
of the $z$-component of the hydrodynamic simulation introduced above,
computed with a French-hat filter with $v=2.3$. The three broken lines
show the $\Delta$-variance spectra computed for the three different
orthogonal projections of the velocity data cube. The solid curve shows
the spectrum computed for the 3-D structure but
rescale as if computed in 2-D by a factor proportional to the lag, to 
compensate for the different exponents of the $\Delta$-variance spectra
depending on the dimensionality of the considered space, and shifted 
by a factor $\pi/4$, to account for the average reduction of a
random structure size when projected from 3-D onto a plane. The solid
line, thus represents our best knowledge on the velocity structure
actually present in the simulations, while the broken lines represent
possible observer's views onto that structure.

We find a considerable variation between the $\Delta$-variance spectra
seen in the different directions, giving a feeling for the statistical
uncertainty when measuring the scaling in turbulent 
simulations\footnote{The computational uncertainty given by the finite
size of any data set was discussed in detail by \citet{Bensch}.}. 
Nevertheless, both the peak position and the exponent at small lags
agree between all three curves. Compared to the full 3-D
structure, there is, however, a systematic shift of the peak to
somewhat larger lags and a slight reduction of the slope at small lags.
For the 3-D structure the peak is seen at a lag of about
0.075, which is about 20\,\% smaller than the expected scale for the
maximum variation, while the peak for the projections falls
between 0.081 and 0.092, i.e. only 10\,\% below the expected scale.
In the 2-D projections, the peak is always broader and the slope
at small lags is always somewhat shallower, similar to the impact
of broader filter functions. It is important to notice, that the
corresponding plots for the turbulent density structure, showing
a shallower scaling at small lags, exhibit a very good match between
the $\Delta$-variance spectra computed in 2-D and in 3-D, i.e. neither
a shift of the peak nor different slopes at small lags.

This seems to indicate that the turbulent velocity cascade does
not behave fully isotropic. A similar effect would be expected for
grid-based hydrodynamic simulations where the dissipation is strong 
between neighbouring cells in the $x$-, $y$-, and $z$-directions.
However, the simulations studied here used an SPH code which should
not show any intrinsic anisotropy. The cause for the anisotropy of
the turbulent velocity scaling is thus so far unknown.

\subsection{Maps with variable noise}
\label{sect_andre}

Maps with variable noise are obtained e.g.
in observations with detector arrays showing a pixel-to-pixel
variation in the sensitivity. They are produced {\changed 
in single-pixel observations when a drift in the receiver sensitivity
or the atmospheric conditions changes the noise in the data during
the measurement and they result from mosaicing observations with 
variable integration times for different regions of the field.
All these cases can be analysed in terms of the improved 
$\Delta$-variance} as long as
the spatial distribution of the noise across the map is well known
so that a corresponding significance map can be defined which
is used to weight the different points in the $\Delta$-variance analysis.
Then the
$\Delta$-variance spectrum is able to distinguish between small-scale
noise contributions in regions with a high noise level and real
small-scale structures in regions with a low noise level.

{\changed As a challenging example of a data set where a variable noise
is produced by the observation of different points of the map
with varying integration time we use the 1.3~mm continuum map of
$\rho$ Oph obtained by \citet{Motte}.
It is the result of a mosaicing observation where an efficient use of
the array receiver is given by several observations of the source
with different orientations of the array.} The combination of
these observations then results in a poorer coverage of the outer
regions of the source compared to the central regions 
in terms of the total integration time spent on each point.
If the source is covered in total with $N\sub{tot}$ observations,
we can characterise the integration time at each point by the
number of coverages including this point $1 \le N \le N\sub{max}$,
where $N\sub{max} \le N\sub{tot}$.
As the noise at each point is inversely proportional to the
square root of the integration time, we can use the value
of $\sqrt{N}$ as a measure for the data reliability across
the map. 

\begin{figure}
\epsfig{file=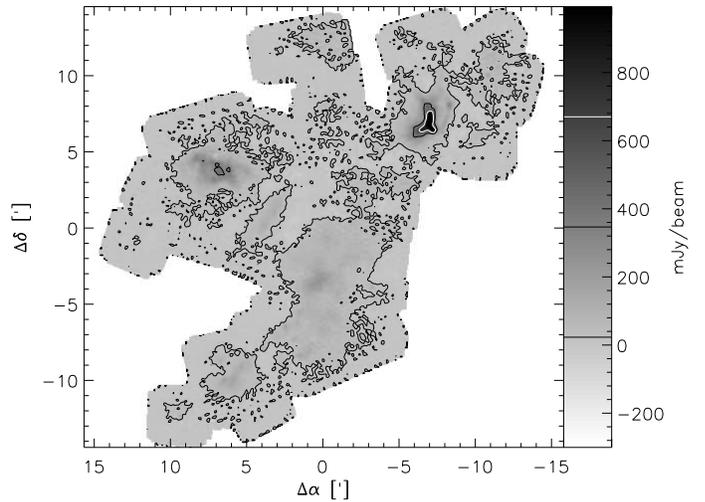, angle=90, width=\columnwidth}
\caption{1.3~mm continuum map of $\rho$ Oph taken by \citet{Motte}.
The intensity is normalised to a 15$''$ beam. In contrast to
\citeauthor{Motte} we plot the intensity on a linear scale because
the structure analysis by means of the $\Delta$-variance only measures
this linear behaviour. For a better contrast of the picture
the intensity scale is truncated here at 1000~mJy/beam whereas
the map contains a few points with intensities up to 1700~mJy/beam
in the brightest core.} 
\label{fig_ophmap}
\end{figure}
\begin{figure}
\epsfig{file=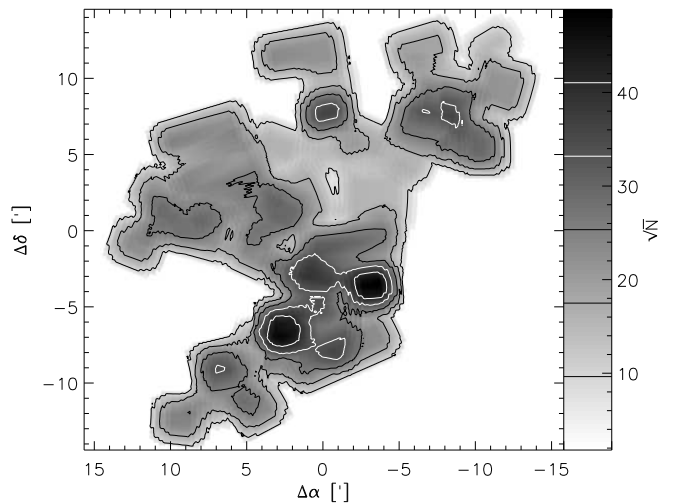, angle=90, width=86mm}
\caption{Map of weights for the intensity map from Fig. \ref{fig_ophmap}
given by the square root of the number of
integrations at each point.}
\label{fig_ophweights}
\end{figure}

{\changed In addition to the variable noise the map has highly irregular boundaries.}
Figure \ref{fig_ophmap} shows the intensity map. In contrast to
the original publication we show the intensity with equidistant
contours on a linear scale
because the linear-scale presentation gives a better feeling for the structure
that is measurable {\changed by means of a statistical method like the $\Delta$-variance analysis.}
To emphasise the irregular noise behaviour we plot one contour at 
20~mJy/15$''$-beam, which is below the noise level in the outer
parts of the map and above the noise level in the inner parts. Consequently,
this contour shows
partly real structure and partly artificial structure from the noise.
Figure \ref{fig_ophweights} contains the corresponding map of 
significance values defined as the square root of
the number of integrations at each point, $\sqrt{N}$, thus measuring
the inverse noise RMS.

\begin{figure}
\epsfig{file=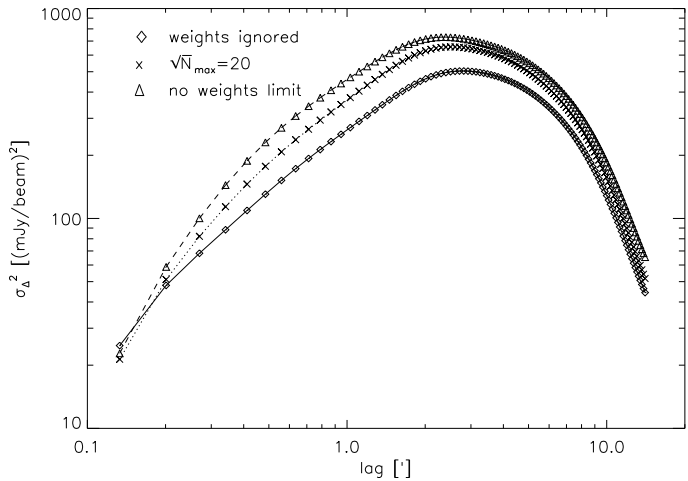, angle=90, width=\columnwidth}
\caption{$\Delta$-variance spectra of the 1.3~mm $\rho$ Oph map
computed using three different weighting functions. Either all observational
weights are ignored, they are completely taken into account, or
weights above $\sqrt{N}=20$ are set to this maximum level.}
\label{fig_ophplot}
\end{figure}

To demonstrate the influence of the significance weighting we show in Fig.
\ref{fig_ophplot} the $\Delta$-variance spectra computed for the
$\rho$ Oph map using three different weighting functions. The lower spectrum
is generated when the weights are ignored, i.e. simply
set to unity 
at all valid data points. The upper graph is produced when
the full weighting function from Fig. \ref{fig_ophweights} is used.
The intermediate curve follows when we introduce and upper limit to
the weighting function motivated by the idea that above a certain
significance limit a further reduction of the noise level does
not improve the structure characterisation any more.
All computations use the filter truncation outside of the area
where data have been taken. The irregular boundaries make it impossible
to construct a useful periodic continuation here. 

\begin{figure}
\epsfig{file=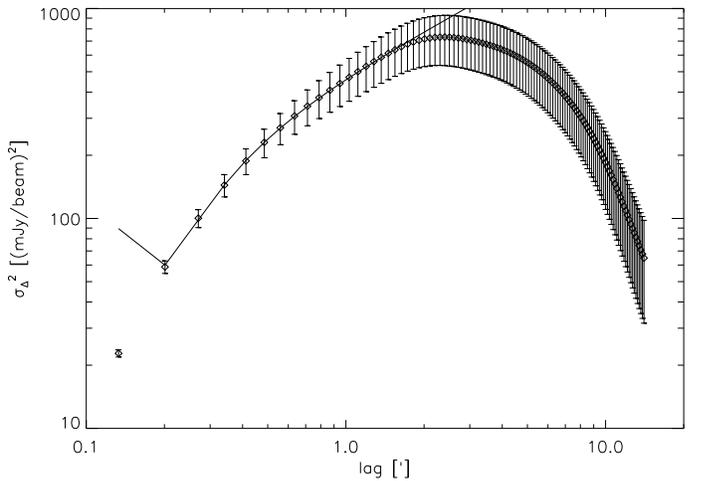, angle=90, width=\columnwidth}
\caption{$\Delta$-variance spectrum for the $\rho$ Oph continuum map
with the full weighting function plotted with error bars indicating
the statistical uncertainty of the measurement. 
The connecting line
represents the power-law fit to the spectrum including the beam
convolution effect. The point at the smallest lag cannot be fitted
by the first order approximation of the beam shape (see Bensch
et al. 2001).}
\label{fig_ophdelta}
\end{figure}

The $\Delta$-variance spectrum computed without weights seems to
indicate a wide range of scales with a power-law behaviour from
about 0.2$'$ to 2$'$ whereas the $\Delta$-variance spectrum computed
with the full weights shows a steepening {\changed starting at 0.5-0.7$'$}. \citet{Bensch}
have shown, however, that such a behaviour is exactly to be expected
from the finite beam of the observations. The data are given
at a {\changed resolution of 15$''$ and the corresponding beam smearing
is known to steepen the $\Delta$-variance spectrum up to scales of about
one arcminute.} This steepening can be modelled theoretically  using the
analytic expressions for the beam convolution from \citeauthor{Bensch}.
In fact, the $\Delta$-variance spectrum computed from the $\rho$ Oph
map with full weights can be fitted by a single power-law 
structure with $\alpha=0.68$ {\changed from $\la 0.2'$ up to about 2$'$
and a convolution function of a 15$''$ HPBW beam} 
(see Fig. \ref{fig_ophdelta}).
The fitted exponent of 0.68 falls into the range measured in molecular
line observations of molecular clouds covering exponents between
0.5 and 1.3 \citep{Bensch, Elmegreen, Falgarone04}.
In contrast, the two lower curves cannot be fitted in the same
way. These spectra would
lead to the conclusion of a surplus of small-scale structure
relative to a power-law scaling relation. 
Such a relative surplus of structure on small scales is hard
to explain {\changed as it would require additional driving processes on 
these scales overcompensating the known dissipation of turbulence 
by ambipolar diffusion and molecular viscosity \citep{Klessen}. 
Gravitational collapse is not able to create these structures;
it always affects the whole $\Delta$-variance spectrum,
not only the small-scale tail \citep{OKH}. A surplus of small-scale 
structure is also in contrast to the analysis of \citet{Motte}
who found a relative lack of small structures in terms of a flatter
clump mass spectrum for small clumps.}

Thus we conclude from the scaling
behaviour that the full weighting of intensity maps by their inverse
noise RMS results in the most reliable $\Delta$-variance spectra.
With this weighting the $\Delta$-variance analysis is able to 
distinguish insignificant small-scale structure, dominating the lowest contour
in Fig. \ref{fig_ophmap}, from significant structures which are intuitively better
presented by the contours chosen in the original plot by
\citeauthor{Motte}.
The increase of the absolute value of the $\Delta$-variance at
large lags when using the
weighting function is explained by the relative increase of the
contribution of the bright cores in the map when virtually reducing the
map size by weighting the outer parts by lower significance values.

To get a feeling for the reliability of the different points
in the $\Delta$-variance spectrum we plot in Fig. \ref{fig_ophdelta}
the $\Delta$-variance spectrum 
including the error bars. The error bars arise from the
statistical uncertainty of the measurement of the average
variance in a filtered map (Bensch et al. 2001).
Due to the lower number of statistically independent points 
in maps convolved with a larger filter, the $\Delta$-variance
is most uncertain at the largest lags. In spite of
the large error bars, the general scaling behaviour can be
accurately traced.
The solid line shows the fit to the data using a power-law 
description of the structure scaling and the convolution by
a 15$''$ beam.

In the $\Delta$-variance spectrum one can clearly see that
the dominating structure has a scale of about 2.2$'$, i.e. 0.1~pc.
This corresponds to the typical size of the cores
identified by \citeauthor{Motte}. The contribution of significant
structures is continued up to 7-9$'$, i.e. 0.3-0.4~pc. 
{\changed This scale agrees with the size of the
largest identified core. Above about 9$'$ the $\Delta$-variance
spectrum decays with $\alpha=-2$ indicating a lack of further correlated
structure on larger scales.} It is
not clear whether the subtraction of large-scale emission
unavoidable in the used observing mode has
removed some large-scale correlation which is present in the
cloud but cannot be detected from the map.

At small lags the spectrum indicates no separation between the
scales where ``cores'' and ``clumps'' (condensations) were
defined by \citeauthor{Motte}. {\changed The fit in Fig. \ref{fig_ophdelta}
shows that on scales from 0.2$'$ to 2$'$, i.e. within a dynamic range
of a factor ten, the spectrum is described by a single power-law
smoothed by the observational beam. The $\Delta$-variance spectrum 
suggests} that the same
processes drive the formation of the somewhat larger ``cores'' and
the somewhat smaller condensations. {\changed The break in
the spectrum at 0.4$'$, that seems to suggest a change in the
scaling law of the observed structure, is only produced by
the beam smearing and is quantitatively in agreement with a 
continuation of the power law observed on larger scales down to 
at least 0.2$'$.
The underlying structure can be described by} a perfect a power law
in contradiction to the clump mass spectrum studied
by \citeauthor{Motte} who found a significant turn-down at a mass
of about 0.5 $M_{\sun}$. This difference is even more intriguing
because of the opposite situation in molecular line studies of the
Polaris Flare where the clump mass spectrum shows a perfect power law 
\citep{Heithausen98} but the $\Delta$-variance spectrum shows
a steepening towards small scales (Bensch et al. 2001, see also
Ossenkopf et al. 2000). 
From the theoretical modelling of the translation of a clump spectrum
into a corresponding power spectral index of an fBm by \citet{Stutzki}, we
would expect a {\changed fixed relation between the measured clump mass spectrum
and the corresponding $\Delta$-variance spectrum in both cases.
However, our examples violate this relation.
The mass-size relation of the clumps may 
be affected by optical depth effects and a large part of the observationally
identified clumps may result from the superposition of different
structures along the line of sight, not well separated in velocity
space \citep{Ballesteros-Paredes, LaSerena}. Further systematic studies
are necessary} to understand the actual physical processes interrelating
the structure size spectra and the clump mass spectra.

{\changed We can compare the new results with the outcome of previous 
$\Delta$-variance analyses, because all previous conclusions on
the slopes of the $\Delta$-variance spectra remain valid. The new 
$\Delta$-variance method has improved our ability to precisely detect
prominent scales, it has calibrated the absolute scales and it increased the
statistical significance of the spectra by taking variable data 
reliability and edge effects into account, but none of these points
should significantly affect the general scaling behaviour measured in 
our previous papers. The $\rho$ Oph map shows a behaviour which is
intermediate between that observed e.g. by \citet{Bensch} in
molecular molecular lines, where we find a power-law $\Delta$-variance
spectrum on small scales and a dominance of large-scale structure,
and the spectrum measured for the 1.3~mm continuum map of Serpens 
\citep{Testi} analysed by \citet{OKH}, where small cores dominate
the spectrum resulting in a steep decay} on large scales.
In the $\rho$-Oph map we find both effects in one spectrum. The dense cores
represent the dominating size scale but we can clearly resolve
the scaling of significant structure on smaller scales. 

Taking all facts from the scaling behaviour and the clump
size detection together we conclude that an appropriate characterisation
of the map in terms of a $\Delta$-variance spectrum is only
obtained when weighting the intensity maps by the inverse
noise RMS.
Otherwise a variable noise in the spectra always tends to mimic
small-scale structure which might be taken for real.

\subsection{Velocity centroid maps}

The situation is more difficult in the analysis of maps representing
other quantities than intensities. Then the weighting
function given by the inverse noise RMS of the observation is not
necessarily a good measure for the significance of the data. We
study one such example here.

Investigating the velocity structure in the Polaris Flare molecular
cloud, \citet{OML} applied the ordinary $\Delta$-variance analysis to maps of
centroid velocities in CO line data. The scaling behaviour
of the velocity field measured in terms of the $\Delta$-variance
spectrum was compared to model simulations of interstellar
turbulence. The data sets were given by three nested
CO maps taken with different telescopes at different resolutions.
The maps taken at high resolutions with the IRAM 30~m telescope
\citep{Falgarone} and the KOSMA 3~m telescope (Bensch et al. 2001) only
covered regions with sufficiently bright emission so that 
a reliable determination of the line centroid velocities
was possible at all points.  The centroid $\Delta$-variance 
spectra derived for these two maps showed a continuous power-law
spectrum with a slight steepening towards the
smallest lags.

\begin{figure}
\centering
\epsfig{file=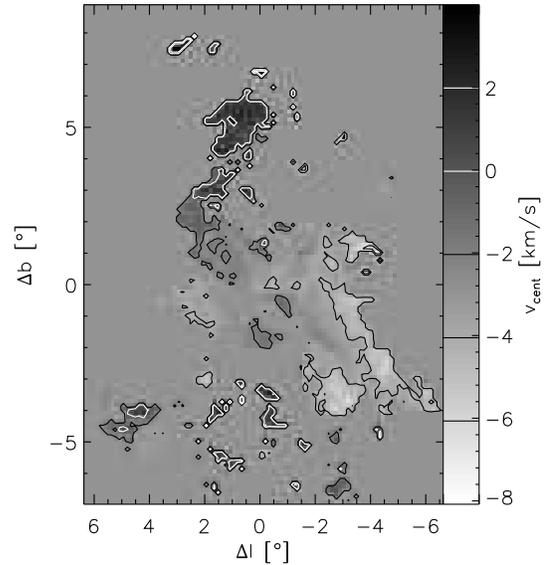, angle=90, width=7cm}
\caption{Map of velocity centroids measured in the CO 1-0 map
of the Polaris Flare taken with the CfA 1.4~m telescope \citep{Heithausen90}.
For all points with an integrated intensity below 0.65 K km/s
a reasonable determination of the centroid velocity was impossible
so that the average velocity of the cloud {\changed (-2.97 km/s)} was assigned there.
The axes are labelled relative to the zero position of $l=
123^{\circ}\!\!.628$, $b=24^{\circ}\!\!.93$ in Galactic coordinates.}
\label{fig_polariscmap}
\end{figure}
\begin{figure}
\centering
\epsfig{file=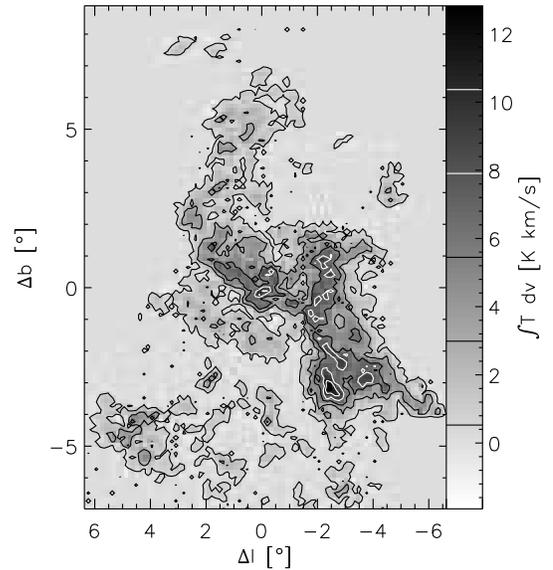, angle=90, width=7cm}
\caption{CO 1-0 intensity map corresponding to the velocity
centroid map in Fig. \ref{fig_polariscmap}. The intensities
provide a measure for the significance of the centroid velocities.}
\label{fig_polarismap}
\end{figure}

In contrast, the map on the largest scale taken with the CfA 1.2~m telescope
\citep{Heithausen90} contains many data points where no emission
above the noise limit was detected. Moreover, it was difficult to
obtain a reliable determination of the centroid velocities in
regions where the line intensities only exceeded the noise RMS by
a factor of a few (see Ossenkopf \& Mac Low 2002). The resulting
$\Delta$-variance spectrum did not show a continuation of 
the power-law behaviour from the two maps on smaller scales, but 
turned essentially flat. This is in contradiction to an eye inspection
of the centroid map plotted in Fig. \ref{fig_polariscmap} showing
a large-scale velocity gradient which should appear as well as
large-scale structure in the $\Delta$-variance spectrum.
 The corresponding map of line integrated intensities,
plotted in Fig. \ref{fig_polarismap}, shows that the map
contains large regions without emission. When the ordinary
$\Delta$-variance counts their centroid velocity with the same
weight as that from points in the actual molecular cloud
the ``empty regions'' statistically hide the variations in the
regions with significant values. {\changed The virtual lack of large-scale
velocity variations in the $\Delta$-variance spectrum is thus
due to the missing significance weighting.} 

\begin{figure}
\epsfig{file=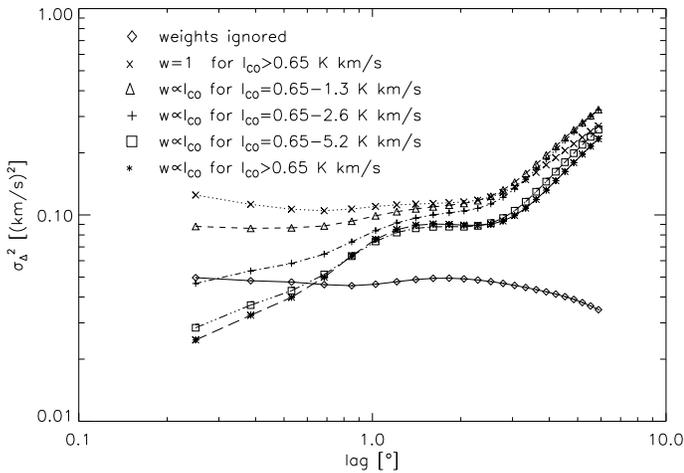, angle=90, width=\columnwidth}
\caption{$\Delta$-variance spectra of the Polaris Flare velocity centroids
computed with different weighting functions characterising the
significance of the velocities at each point.}
\label{fig_polarisplot}
\end{figure}

{\changed
The knowledge on the 
reliability of the centroid velocities coming from the
corresponding line intensities has to be taken into account
using the improved $\Delta$-variance analysis.} Unfortunately,
it is not obvious how the significance of the centroid velocities
is related to the line intensities. We have tested weighting functions
based on three assumptions: {\bf i)} The significance of the centroid
velocities is determined by the integrated line intensities at each point.
{\bf ii)} The zero value of the weighting function corresponds to a minimum
intensity of 0.65~K km/s which means that at least 10 velocity channels
show an intensity above the noise RMS. 
{\bf iii)} Integrated intensities above some limit
do not further increase the significance of the centroid velocities.
The weighting function is unity at all points with higher intensities.

Figure \ref{fig_polarisplot} shows the resulting $\Delta$-variance spectra
of the centroid map when weighting functions with different
upper intensity thresholds are applied. For an easy comparison to the
previous results we also include the graph obtained without any weighting. 
The introduction of the weighting
function results in a completely different $\Delta$-variance
spectrum at large lags. Whereas the previous computation showed
a flat spectrum we find now a strong increase of the spectrum 
above $3\degree$. Unfortunately, the results only give rough guidelines
for the selection of the optimum weighting function. {\changed The best continuation
of the $\Delta$-variance spectra from the smaller scales provided
by the KOSMA map is obtained when the upper intensity limit 
corresponds to four times the lower limit.} The two
curves only using a narrow dynamic intensity range for the weighting
function are still heavily influenced by observational noise visible
as increased $\Delta$-variance values at small lags. On the other
hand it is not clear whether the complete suppression of the noise
effect in the two curves with the highest upper limits is
realistic, so that
we conclude that in this example the significance of the centroid
velocities does not further increase for integrated intensities above
2.6 K km/s. Nevertheless, a final answer to this question still
has to come from a
theoretical model for the quantitative impact of the observational
noise on the noise in centroid velocities.

In spite of the uncertainty of the noise contribution at small
lags, we can draw new essential conclusions on the velocity 
structure of the Polaris Flare from the $\Delta$-variance spectrum.
The power-law scaling behaviour of the velocity structure
detected previously only for smaller sizes is now 
continued up to scales of 1.3$\degree$, i.e. 3-4~pc. A single
power law with an exponent $\alpha\approx 0.9$ can be used to cover
the length scale range of about a factor 100. The $\Delta$-variance
spectrum also shows a plateau around 2$\degree$, i.e.
5~pc, indicating the relative deficiency of motions on that scale.
Above $3\degree$ the $\Delta$-variance spectrum rises again
tracing the global velocity gradient visible in Fig. \ref{fig_polariscmap}.
This behaviour indicates that the large-scale gradient
is not converted directly by shear motions into the turbulent
cascade but that the turbulent cascade starts at somewhat smaller
scales. This points towards a shock producing the large-scale gradient
which is only converted into turbulent energy on the scale of 
previously existing density fluctuations in the cloud and excludes
Galactic rotation as the main driving force.

\section{Conclusions}

{\changed In paper I we have proposed two essential improvements of the 
$\Delta$-variance analysis. Here, we have tested their actual
impact when applied to data sets characterising interstellar 
turbulence.

The first improvement was the introduction of a weighting function
for each pixel in the map. This allows us to study data sets with a
variable data reliability across the map and to simultaneously 
solve boundary problems even for maps with irregular boundaries.}
Maps with a variable data reliability are eventually obtained in
most observations, either due to a local or a temporal variability
of the detector sensitivity or the atmosphere or due to 
different integration times spent for different points of a map.
By applying the improved $\Delta$-variance analysis to observed data
we find that only the use of a significance function to weight
the different data points allows us to distinguish the influence
of variable noise from actual small-scale structure in the maps.
In the analysis of intensity maps the 
weighting function is best provided by the inverse RMS in the
data points. The situation is more complex for derived
quantities, {\changed like centroid velocities, without a simple analytic
relation between the uncertainty of the quantity and the
observational noise.} Here, in general two thresholds can be defined --
a lower threshold below which all data have to be ignored and
an upper threshold above which the significance of the data
is not further improved by decreasing the noise. For the
centroid velocities this means that the integrated line intensities
between the two boundaries may serve as weighting function.

The second improvement of the $\Delta$-variance analysis
is its optimisation with respect to the shape of the wavelet used
to filter the observed maps. {\changed The application of different 
filters in the analysis of hydrodynamic simulations confirmed
the result from paper I, that a Mexican-hat filter with a diameter
ratio $v=1.5$ is well suited to resolve prominent structure scales
and to measure the slope of the turbulent cascade, however,
it turned out that the impact of the detailed shape of the
$\Delta$-variance filter is less significant for realistic data
than for the artificial test data used in paper I. The 
turbulent structure is well resolved for a wide set of
filters as long as one consistent filter shape is used throughout
the full analysis of a data set.
Comparing the density and velocity structure of the simulations
shows a small but significant shift between the scale of the most
prominent velocity structures, created by the energy injection, and
the most prominent density structures produced by the velocity field.}

Applying the new method to the example of the dust emission map of
$\rho$ Oph by \citet{Motte} shows that the spatial scaling behaviour there
can be described perfectly by a power law interconnecting the range of
small clumps and more massive cores. The method can reproduce
the size of the dominant cores and we find no indication for large-scale
correlation between the clumps and cores in the data.
The $\Delta$-variance spectrum shows no break in the scaling behaviour
between cores and condensations in contrast to the mass spectrum
derived by \citet{Motte}. The reason for the different behaviour
of the two measures has to be topic of a future investigation.

In the example of the analysis of the velocity structure in the
Polaris Flare we show that the power-law scaling behaviour
established by \citet{OML} for the small scales is continued to
large scales. However, a plateau in the $\Delta$-variance spectrum
around 5~pc indicates that the existing large-scale velocity gradient
is not converted directly into a turbulent
cascade. A possible explanation for this behaviour is the
existence of a shock producing the large-scale gradient which
is only converted into turbulent energy on the scale of the
individual density fluctuations in the cloud. This scenario
would be consistent with the affiliation of the molecular cloud
to a large H {\sc I} supershell by \citet{Meyerdiercks}.

{\changed
Combining the results from the turbulence simulation with the
analysis of the Polaris Flare velocity structure indicates that
a large-scale velocity field does not automatically produce
density structures on those scales, but that a full turbulence
cascade covering density and velocity fluctuations evolves
predominantly at seeds of primordially existing density 
fluctuations, which may have been produced by previous 
velocity fields on larger scales. When interpreting turbulent
structures in interstellar clouds it has to be taken into account
that close to the scale of the energy injection a statistical
analysis of the turbulent cascade is always affected by low 
number statistics as few density ``seeds'' may dominate the 
shape of the scaling relations there. A reliable statistics
is only given on smaller scales.} 

\begin{acknowledgements}
We thank R.~Klessen for providing us with the numerical data of the
SPH simulations used in Sect.~\ref{sect_mhd}.
We thank Ph.~Andr\'e for the data of the $\rho$ Oph
continuum observations used in Sect. \ref{sect_andre}.
We thank F.~Bensch for useful discussions and J.~Ballesteros-Paredes
for carefully refereeing this paper suggesting significant
improvements. This work has been supported by the Deut\-sche
For\-schungs\-ge\-mein\-schaft through grant 494B.
It has made use of NASA's Astrophysics Data System Abstract Service.
\end{acknowledgements}

\end{document}